\documentclass[pra,twocolumn,showpacs,aps,floatfix,amsmath,amssymb,amsfonts]{revtex4-1}

\usepackage{amsmath}
\usepackage{graphicx}
\usepackage{color}

\graphicspath{{figures/}}

\newcommand{\beq}{\begin{equation}}
\newcommand{\eeq}{\end{equation}}

\begin{document}

\title{Ultracold atoms in quasi-1D traps: a step beyond the Lieb-Liniger model}
\author{Krzysztof Jachymski$^1$}
\author{Florian Meinert$^2$}
\author{Hagar Veksler$^3$}
\author{Paul S. Julienne$^4$}
\author{Shmuel Fishman$^3$}
\affiliation{
$^1$ Institute for Theoretical Physics III, University of Stuttgart, Pfaffenwaldring 57, 70569 Stuttgart, Germany\\
$^2$ 5. Physikalisches Institut and Center for Integrated Quantum Science and Technology (IQST), University of Stuttgart, Pfaffenwaldring 57, 70569 Stuttgart, Germany\\
$^3$ Physics Department, Technion - Israel Institute of Technology, Haifa 3200, Israel\\
$^4$ Joint Quantum Institute, University of Maryland and National Institute of Standards and Technology, College Park, Maryland 20742, USA
}
\date{\today}

\begin{abstract}
Ultracold atoms placed in a tight cigar-shaped trap are usually described in terms of the Lieb-Liniger model. We study the extensions of this model which arise when van der Waals interaction between atoms is taken into account. We find that the corrections induced by the finite range of interactions can become especially important in the vicinity of narrow Feshbach resonances and suggest realistic schemes of their experimental detection. The interplay of confinement and interactions can lead to effective transparency where the one-dimensional interactions are weak in a wide range of parameters.
\end{abstract}

\maketitle

\section{Introduction}
Interacting quantum systems are the subject of great general interest, but they are very hard to treat analytically. The Lieb-Liniger model~\cite{Lieb1963} is one of the rare examples of the exactly solvable case. It describes $N$ identical bosons on a ring of circumference $L$ interacting via the delta function potential 
\beq
V(x)=g_{1D}\delta(x),
\label{delta}
\eeq
where $x$ is the interparticle distance. In the limit $g_{1D}\to\infty$ the problem can be mapped onto noninteracting fermions and the system forms the Tonks-Girardeau gas~\cite{Tonks1936,Girardeau1960,Cazalilla2011}. This phenomenon has been observed experimentally using a tightly confined gas of ultracold neutral atoms~\cite{Kinoshita2004,Paredes2004}. This was possible due to excellent experimental control of the external trapping potential as well as interatomic interactions, which is a unique feature of cold atomic systems~\cite{Bloch2008,Chin2010}. Further studies of quasi-one-dimensional (Q1D) bosons included e.g. measurements of two- and three-body correlation functions~\cite{Kinoshita2005,Haller2011}, creation of a metastable, strongly correlated gas with attractive interactions~\cite{Haller2009}, studying the gas out of equilibrium~\cite{Gring2012,Trotzky2012}, probing the excitation spectrum of the system~\cite{Meinert2015}, and investigating impurity dynamics~\cite{Catani2012,Meinert2016}.

For the Lieb-Liniger model an exact solution has been presented and its properties could be studied in the thermodynamic limit. This was possible due to the simple form of the interaction. The purpose of the present paper is to study situations in which the assumption of point-like particles is not correct. With this motivation in mind, some of us~\cite{Veksler2014,Veksler2016} introduced the modified Lieb-Liniger model in which the interaction potential~\eqref{delta} is replaced by a more complex term
\beq
V(x)=c_0\delta(x)+c_\ell\left(\delta(x-\ell)+\delta(x+\ell)\right),
\label{gll}
\eeq
which aims to account for the finite range of the realistic interactions between atoms. Here, we will study what are the corrections to the simple model~\eqref{delta} for the system of ultracold atoms placed in a Q1D trap and to what extent their properties can be captured by the model~\eqref{gll}. We find that while the extension of the form proposed in~\eqref{gll} describes the properties of cold atoms only for relatively weak interactions, finite range corrections can indeed become crucially important. In general, the first-order correction results in the effective interaction of the form~\cite{Collin2007}
\beq
V(p)=g_{1D}\left(1+g^\prime p^2\right),
\label{truepot}
\eeq
where $\hbar p$ is the 1D momentum. Formula~\eqref{gll} can then be interpreted as a discretized version of~\eqref{truepot}.

The two-body problem in the presence of quasi-1D confinement has been the subject of experimental and theoretical interest for many years~\cite{Olshanii1998,Bergeman2003,Bolda2003,Idziaszek2006,Moritz2005,Naidon2007,Yurovsky2007,Yurovsky2008,Haller2010,Giannakeas2012,Sala2012,Sala2013,Hess2014,Hess2015}. It has been discovered~\cite{Olshanii1998} that the effective 1D interaction strength is strongly affected by the presence of the external trap. Surprisingly, the transmission coefficient can become zero at finite 3D interaction strength, meaning that the 1D interaction becomes divergent although the 3D one does not. This phenomenon is called confinement-induced resonance (CIR) and can be studied experimentally by tuning the 3D scattering length of the atoms using Feshbach resonances~\cite{Chin2010}.

Most previous works examined the case for which the finite energy corrections do not significantly affect the properties of the CIR. This is the case for relatively weak transverse confinement and open-channel dominated Feshbach resonances. Here we show that for closed-channel dominated resonances the scattering properties of the atoms and consequently the effective pseudopotential describing their many-body properties can become strongly energy-dependent. We provide a simple framework which allows for an analytic description of the corrections in terms of the effective range~\cite{Naidon2007}. Focusing on the case of Cs atoms, which feature a number of both broad and narrow Feshbach resonances, we discuss how to demonstrate the role of energy dependence using readily accessible experimental setups.  

This work is organized as follows. In Section II we review the properties of the generalized Lieb-Liniger model introduced in~\cite{Veksler2016}. In Section III we provide the solution to the two-particle problem in the presence of transverse confinement using an energy-dependent pseudopotential. Section IV discusses the corrections to the properties of CIR resulting from energy dependence and the role of effective range. Section V provides the comparison of the realistic atomic scattering results to the predictions of the generalized Lieb-Liniger model and studies the limits of its applicability. The results are discussed and future directions are proposed in Section VI.

\section{Generalized Lieb-Liniger model}
Let us consider a gas of atoms interacting via the three-delta potential~\eqref{gll}. We will concentrate on the low-energy and low-density ($N\ell\ll L$) limit. In this case the model is well approximated by a single delta function potential~\eqref{delta} with effective coefficient $c_{\rm eff}$ given by~\cite{Veksler2016}
\beq
c_{\rm eff}=c_0+2c_\ell+\frac{\frac{mc_\ell \ell}{\hbar^2}\left(2c_0+2c_\ell+\frac{mc_0 c_\ell \ell}{\hbar^2}+\frac{mc_0^2 \ell}{2\hbar^2}\right)}{1-\frac{m^2 c_0 c_\ell \ell^2}{2\hbar^4}-\frac{m c_\ell \ell}{\hbar^2}}.
\label{ceff}
\eeq 
A possible choice of the parameters $c_0$, $c_{\ell}$ is $c_0=2g_{1D}$ and $c_{\ell}=-g_{1D}/2$, such that the sum of the three coefficients $c_0+2c_\ell=g_{1D}$.
In the limit $\ell\to 0$ we have $c_{\rm eff}\to c_0+2c_\ell$, which is an intuitive result expected to hold in the long wavelength limit. Surprisingly, eq.~\eqref{ceff} implies that when the parameter $q=m g_{1D}\ell/\hbar^2$ is of the order of unity, the $c_{\rm eff}$ coefficient strongly deviates from the naive expectation and can change its sign (see Fig.~1 of~\cite{Veksler2016}). When $c_{\rm eff}=0$ the particles are effectively noninteracting.

For a realistic system of atoms one can argue that the generalized Lieb-Liniger model (GLL) can provide a first order correction beyond the standard delta-function interaction~\eqref{delta}. In the following, we will study the atomic scattering problem in a harmonic trap and derive the expressions for $g_{1D}$ and $\ell$ in terms of the 3D scattering length, effective range and confinement strength and compare the results with the prediction of formula~\eqref{ceff}.

\section{Atomic scattering in a quasi-1D waveguide}

We now consider an experimentally realistic system of ultracold atoms interacting with a van der Waals potential $V_{\rm vdW}(r)=-C_6/r^6$. The characteristic range of this interaction can be defined as $\bar{a}=\frac{2\pi}{\Gamma(1/4)^2}(2\mu C_6/\hbar^2)^{1/4}$, where $\mu$ is the reduced mass of the atomic pair~\cite{Gribakin}. As $\bar{a}$ is much smaller than typical interatomic distances in the gas and the de Broglie wavelength, and the collision energies are typically in the nK regime such that only $s$-wave interacions are relevant, it is justified to describe the interaction using the Fermi pseudopotential
\beq
V(\mathbf{r}) = \frac{2\pi\hbar^2 a_{3D}}{\mu}\delta(\mathbf{r})\frac{\partial}{\partial r}r,
\label{pseudo0}
\eeq
where $\hbar^2 k^2/(2\mu)$ is the kinetic energy of the relative motion and the essential quantity is the scattering length of the potential $a_{3D}$ related to the phase shift $\delta_{3D}$ via $a_{3D}=-\tan\delta_{3D}(k)/k$. This approach leads to the well-known Fermi-Huang pseudopotential with the coupling constant $g_{3D}=\frac{2\pi\hbar^2 a_{3D}}{\mu}$.

In the presence of an external trap, one should take into account that the scattering phase shift is in fact weakly energy dependent and use the corrected pseudopotential
\beq
V(\mathbf{r}) = -\frac{2\pi\hbar^2}{\mu}\frac{\tan\delta(k)}{k}\delta(\mathbf{r})\frac{\partial}{\partial r}r,
\label{pseudo1}
\eeq
which leads to much better agreement with numerical calculations in which the Schr\"{o}dinger equation was solved with a model potential and compared to the constant pseudopotential ~\eqref{pseudo0} predictions ~\cite{Bolda2002,Blume2002,Idziaszek2006,Naidon2007,Kalas2008,Yu2011}. This is equivalent to replacing the scattering length by its energy-dependent generalization
\beq
a_{3D}(k)=-\frac{\tan\delta_{3D}(k)}{k}.
\eeq
A similar correction can be introduced in free space, where in the mean-field approach one has to include first-order corrections to the scattering amplitude in order to obtain an extended Gross-Pitaevskii equation~\cite{Fu2003,Collin2007}.

We now assume that the atoms are subject to strong radial harmonic confinement $\frac{1}{2}\mu\omega^2\rho^2$ ($z$ is the free direction, while the motion in $\rho=\sqrt{x^2+y^2}$ is restricted) characterized by trapping frequency $\omega$. The center of mass motion in the harmonic trap can be separated out and the relative part of the wave function for two atoms fulfills the Schr\"{o}dinger equation
\beq
\left(\frac{\hbar^2 \nabla^2}{2\mu}+V(\mathbf{r})+\frac{1}{2}\mu\omega
^2 \rho^2\right)\psi(\mathbf{r})=E\psi(\mathbf{r}).
\eeq
The characteristic length scale for the trapping potential can be defined as $d=\sqrt{\hbar/\mu\omega}$. At large separations where the interaction potential can be neglected, the wave function can be decomposed into the incoming and scattered part, following
\beq
\psi\stackrel{r\to\infty}{\longrightarrow} \psi_{nm}(\boldsymbol{\rho})e^{ipz}+\sum_{n^\prime m^\prime}{f^{(+)}_{nm,n^\prime m^\prime}(p)\psi_{n^\prime m^\prime}(\boldsymbol{\rho})e^{ip^\prime |z|}}.
\label{boundcond}
\eeq
Here $f^{(+)}_{nm,n^\prime m^\prime}(p)$ denotes the even part of the 1D scattering amplitude and $\psi_{nm}$ denote the transverse harmonic oscillator states. For identical bosons the odd part of the scattering amplitude vanishes due to symmetry requirements. The total energy $E$ consists of the transverse harmonic oscillator part and the plane wave in the longitudinal direction
\beq
E=\hbar\omega(1+2n+|m|)+\frac{\hbar^2 p^2}{2\mu}=\frac{\hbar^2 k^2}{2\mu}.
\eeq
It is important to note that at short interparticle distances the trapping potential is almost constant and the collision is essentially three-dimensional. Therefore the 1D and 3D momenta ($p$ and $k$, respectively) are not the same even in the ground state of the transverse trap due to the zero-point motion associated with the harmonic potential. Namely, $k^2=p^2+2/d^2$. This can become relevant when $1/d$ cannot be neglected compared to $k$.

By solving the Schr\"{o}dinger equation with the pseudopotential~\eqref{pseudo1} and boundary conditions~\eqref{boundcond}, we obtain~\cite{Bergeman2003,Idziaszek2015}
\beq
\begin{split}
f^{(+)}_{nm,n^\prime m^\prime}(p)=\frac{4\pi a_{3D}(k)}{2ip^\prime}\psi_{n^\prime m^\prime}(0)\psi_{nm}(0)\times \\ \times \left(1+\frac{a_{3D}(k)}{d}\zeta_H\left(\frac{1}{2},\frac{1}{2}-\frac{E}{2\hbar\omega}\right)\right)^{-1},
\end{split}
\label{f1d}
\eeq
where $\zeta_H$ is the Hurwitz zeta function. Due to the properties of harmonic oscillator states, only terms with $m=m^\prime=0$ do not vanish.
We stress that the distribution of the total energy among the transverse and longitudinal degrees of freedom at long range does not affect $a_{3D}(k)$. In the following, we will restrict to the case where only the lowest transverse mode is occupied and denote $f^{(+)}_{00,00}(p)$ as $f(p)$ for simplicity.

The 1D scattering amplitude can be connected to the 1D phase shift $\delta_{1D}$ via~\cite{Olshanii1998,Bergeman2003}
\beq
f(p)=-\frac{1}{1+i\cot\delta_{1D}(p)}
\eeq
and the 1D energy-dependent scattering length can be defined as
\beq
a_{1D}(p)=\frac{1}{p\tan\delta_{1D}(p)}.
\label{a1def}
\eeq
The scattering length obtained this way from~\eqref{f1d} may at first sight appear to be imaginary due to arising $i/p$ term. However, this term is exactly cancelled by the imaginary part of the zeta function, and we get

\beq
a_{1D}(p)=-\frac{d^2}{2 a_{3D}(k)}\left(1-C(k) \frac{a_{3D}(k)}{d}\right).
\label{a1Dp}
\eeq
Here, $C(k)$ is the real part of $=-\zeta_H\left(\frac{1}{2},\frac{1}{2}-\frac{E}{2\hbar\omega}\right)$.
We note that in the $k\to 0$ limit $a_{1D}(p)$ reduces to the well-known expression~\cite{Olshanii1998}
\beq
a_{1D}(p\to 0)=-\frac{d^2}{2 a_{3D}}\left(1-C \frac{a_{3D}}{d}\right)
\label{a1D0}
\eeq
where $C=-\zeta(1/2)\approx 1.46035\ldots$.

Here, we are interested in the leading-order finite energy corrections, for which it is sufficient to use the effective range expansion for the scattering length
\beq
k\cot\delta_{3D}(k)=-\frac{1}{a_{3D}}+\frac{1}{2}r_{3D} k^2+o(k^2),
\label{eqeffrexp}
\eeq
which gives
\beq
a_{3D}(k)\approx \frac{a_{3D}}{1-k^2 r_{3D}a_{3D}/2}.
\eeq
The possible forms of the effective range $r_{3D}$ for realistic van der Waals interactions will be discussed below. 

The effective 1D pseudopotential $V_{1D}=g_{1D}(p)\delta(x)$ contains the coupling strength parameter proportional to the inverse of $a_{1D}$~\cite{Olshanii1998,Bolda2003}
\beq
g_{1D}(p)=-\frac{\hbar^2}{\mu} p \tan\delta_{1D}(p).
\label{g1ddef}
\eeq
When expanding this quantity in power series of $p$ we have to also expand the $C(k)$ term using $\zeta_H(1/2,-x)=i/\sqrt{x}+\zeta(1/2)+\frac{1}{2}\zeta(3/2)x+\ldots$. Up to the second-order term, we then have
\beq
g_{1D}(p)=g_{1D}(1+g^\prime p^2)
\eeq
with
\beq
\frac{\mu}{\hbar^2}g_{1D}=\frac{2}{d}\left(\frac{d}{a_{3D}}-C-\frac{r_{3D}}{d}\right)^{-1}
\label{eqg1d}
\eeq
and
\beq
g^\prime=\frac{d}{2}\frac{r_{3D}-\tilde{C}d}{\frac{d}{a_{3D}}-C-\frac{r_{3D}}{d}},
\label{eqgp}
\eeq
where $\tilde{C}=\zeta(3/2)/8\approx 0.3265\ldots$ comes from the expansion of $C(k)$.

Let us now introduce another important quantity, the transmission coefficient $T$, which describes the part of the flux that goes through the potential barrier in relative coordinates. By definition it is related to scattering amplitude via $T=|1+f|^2$~\cite{Olshanii1998}. It can also  be rewritten in terms of the phase shift as 
\beq
T(p)=\frac{1}{1+\tan^2\delta_{1D}(p)}.
\label{eqtransm}
\eeq
If $T$ reaches unity, this means that $\tan\delta_{1D}$ is zero and the particles are effectively noninteracting.
The position of the CIR can be found by identifying the parameter values at which $\tan^2\delta_{1D}$ diverges and the transmission coefficient becomes zero.

In our derivation we assumed the knowledge about the energy dependence of the scattering length, which for realistic systems has to be obtained by multichannel scattering calculations. It is also possible to formulate the problem in a different way by using quantum-defect theory~\cite{Granger2004,Giannakeas2012,Hess2014,Hess2015,Wang2016}. In this treatment one calculates the short-range 3D $K$~matrix $K_{3D}^{\rm sh}$, which contains all the information about the free-space scattering properties, and then uses frame transformation technique to obtain the 1D scattering quantities via the long-range 1D $K$~matrix $K_{1D}$. This powerful method allows for multichannel, energy-dependent calculations including higher partial waves, which lead to additional narrow resonances.

\section{Role of effective range corrections}
Formulas~\eqref{eqg1d}-\eqref{eqgp} suggest that energy-dependent corrections do not only introduce a $g^\prime$ term, but are also crucially important for determining the value of $g_{1D}$ due to zero-point motion associated with the transverse mode. One immediate consequence of the higher-order terms is that the position of the CIR shifts with respect to the zero-range prediction and becomes energy dependent. %Interestingly, it is even possible to realize the case for which $r_{3D}/d+dr_{3D}p^2/2=-C$, for which the transmission coefficient becomes $T^\star=1/(1+4/p^2 d a_{3D})$ and has a Lorenzian shape with no confinement-induced minimum~\cite{Hess2014}. 
We will now estimate the potential importance of the corrections.

For atoms interacting via a single-channel van der Waals potential the 3D effective range can be calculated analytically and has a universal form depending only on the scattering length~\cite{Gao1998,Blackley2014}
\beq
r_{3D}=\frac{\Gamma\left(1/4\right)^2\bar{a}}{6\pi^2}\left(1-\frac{2\bar{a}}{a_{3D}}+\frac{2 \bar{a}^2}{a_{3D}^2}\right).
\label{reff_univ}
\eeq
Numerical multichannel scattering calculations have verified that this expression holds close to open-channel Feshbach resonances, but can fail on the closed-channel dominated ones~\cite{Blackley2014}. However, an effective formula for the effective range as a function of magnetic field $B$ in the vicinity of the resonance can be given as~\cite{Blackley2014}
\beq
r_{3D}(B)\approx \frac{v+r_0(a_{3D}(B)-a_{ex})^2}{a_{3D}(B)^2},
\label{reff_gen}
\eeq
with $v$, $r_0$ and $a_{ex}$ being fit parameters depending on a particular resonance. From this expression we observe that the effective range diverges at the zero crossing of $a_{3D}$.

Werner and Castin~\cite{Werner2012} proposed an extension of formula~\eqref{reff_univ} to the case of a Feshbach resonance by adding the term 
$-2R_\star\left(1-\frac{a_{bg}}{a_{3D}}\right)^2$ which accounts for the Feshbach resonance. Here $R_\star=\hbar^2/(2\mu a_{bg}\Delta\delta\mu)$, $a_{bg}$ is the background scattering length away from resonance, $\Delta$ is the resonance width and $\delta\mu$ is the magnetic moment difference between the channels. This term shows that $r_{3D}$ can take negative values, and vanishes when the parameter $s=a_{bg}\Delta\delta\mu$ becomes large, which is the case for open-channel dominated resonances~\cite{Chin2010} (a similar result close to the resonance has been derived in~\cite{Jachymski2013}).
We then expect that the effective range corrections can be much more important for narrow resonances since the $v$ and $r_0$ parameters can take very large values.

\begin{figure}
  \centering
  \includegraphics[width=0.5\textwidth]{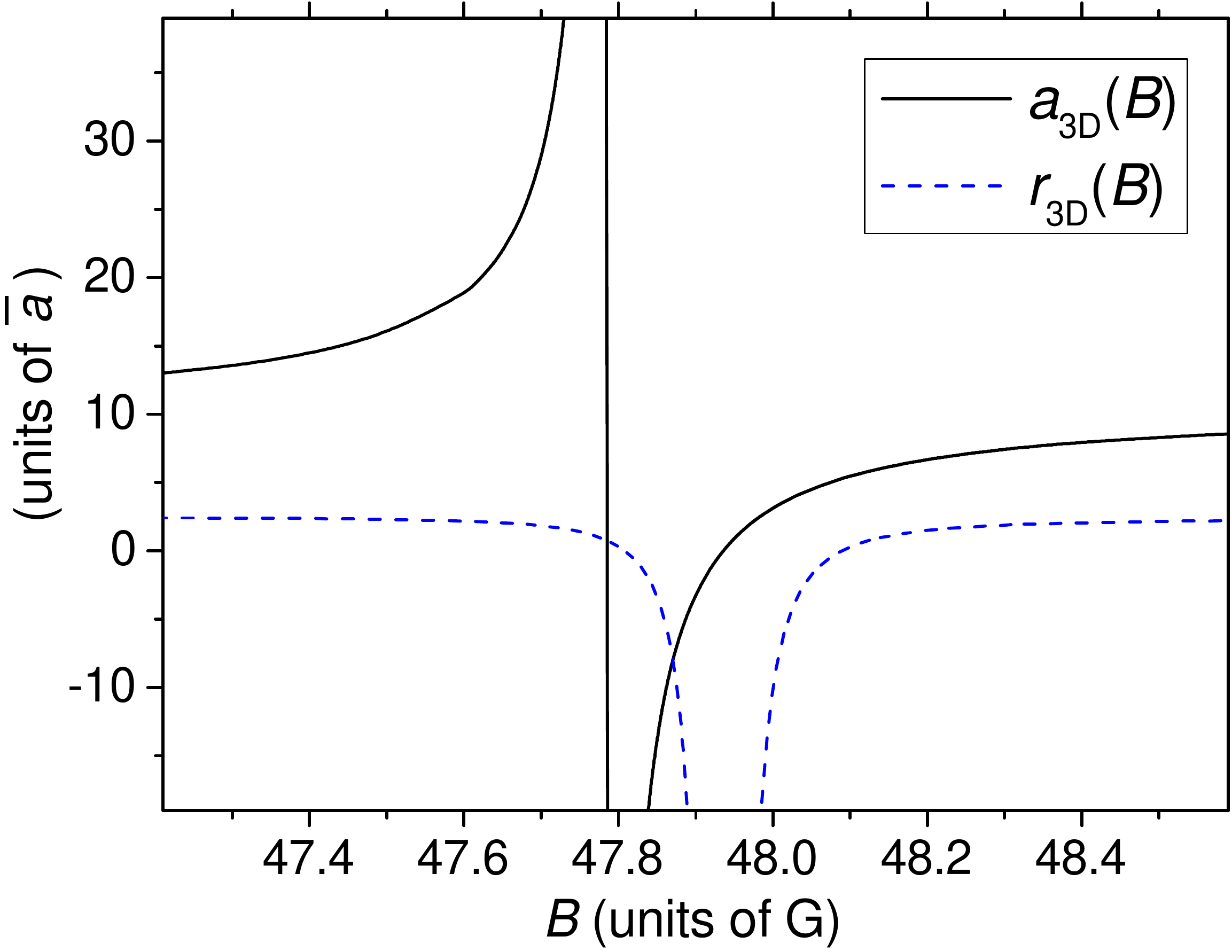}
  \includegraphics[width=0.5\textwidth]{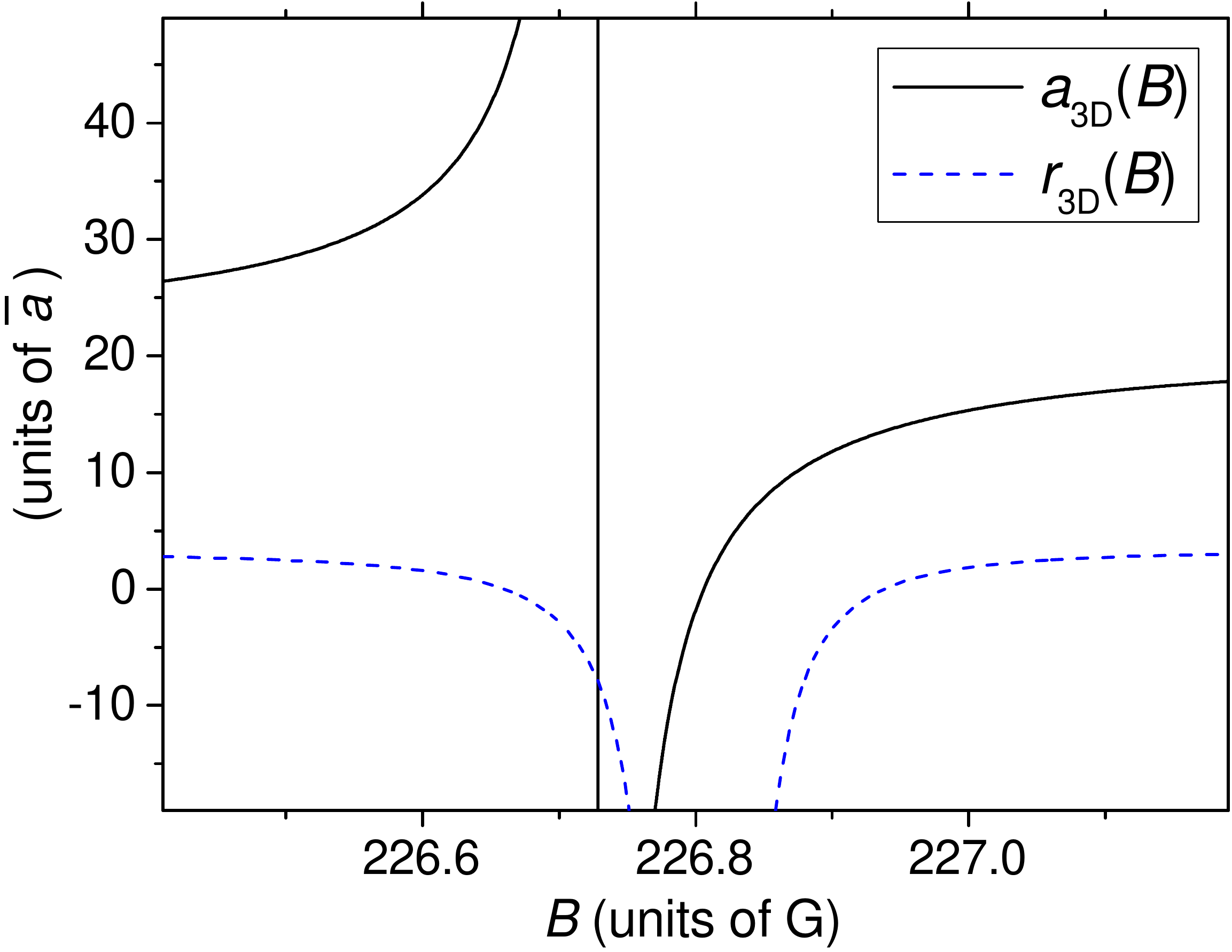}
  \caption{\label{fig:ascreff}3D scattering length $a_{3D}$ (black solid line) and effective range $r_{3D}$ (blue dashed line) in units of $\bar{a}\approx 96\,a_0$ as a function of magnetic field $B$  for two narrow Cs resonances.}
\end{figure}

We will focus on the case of Cs atoms, which have been used by the Innsbruck group to experimentally demonstrate the existence of CIR~\cite{Haller2010} and to study different aspects of 1D many-body systems~\cite{Meinert2015,Meinert2016}. Cs displays a variety of both broad and narrow resonances which have been extensively studied~\cite{Berninger2013}. We will study the case of a narrow resonance situated at a magnetic field of $\sim 47.8$G, and an even more narrow one at~$\sim 226.7$G. The scattering length and effective range for these two cases obtained by fitting formula~\eqref{reff_gen} to the results of multichannel scattering calculations of~\cite{Berninger2013} are shown on Fig.~\ref{fig:ascreff}.

So far in typical experiments the transverse trapping frequency was in the few kHz range. We will assume a typical value $\omega\approx 2\pi \times 14.5$kHz which gives $d\approx 1930\,a_0$. As the mean van der Waals length for Cs $\bar{a}\approx 96\,a_0$, the ratio of the length scales is $d\approx 20\bar{a}$. 

\begin{figure}
  \centering
  \includegraphics[width=0.5\textwidth]{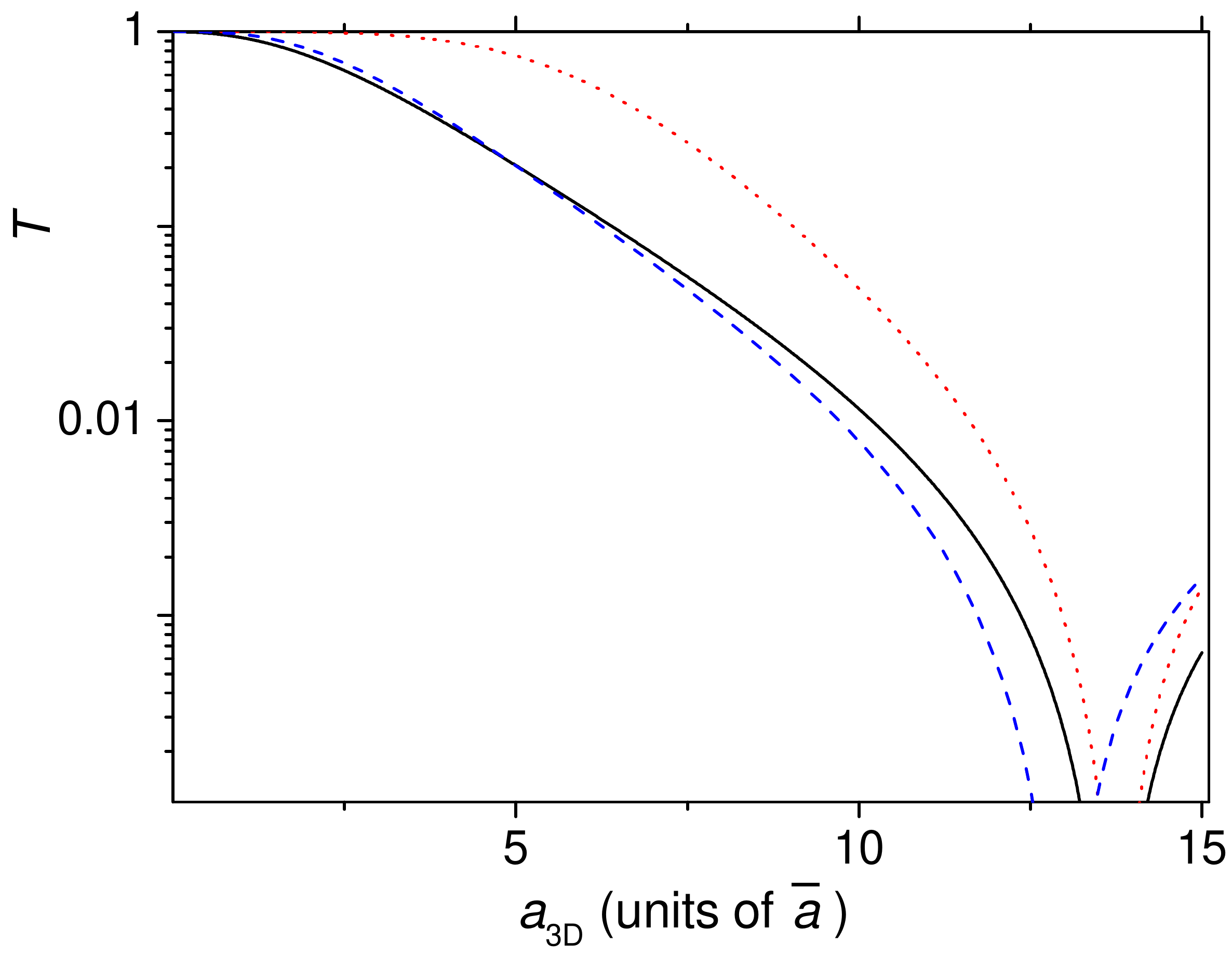}
  \caption{\label{fig:transmwide}Transmission coefficient $T$ as calculated from eq.~\eqref{eqtransm} for different Cs resonances as a functions of scattering length $a_{3D}$ for a wide trap $d=20\bar{a}$ and low energy $p=0.01\bar{a}^{-1}$ (which corresponds to around 15 nK longitudinal kinetic energy). The black line shows the broad resonance case, the blue dashed line depicts the predictions for the $47.8$G resonance and the red dotted line gives the results for the resonance at $226.7$G.}
\end{figure}

Figure~\ref{fig:transmwide} shows the transmission coefficient for different resonances as a function of the scattering length for the $d=20\bar{a}$ case. We observe that the curves corresponding to the single-channel case~\eqref{reff_univ} and to the $47.8$G resonance are not much different. On the other hand, for the $226.7$G resonance the transmission is much closer to unity in a wide range of the scattering length values. This is because the effective range term in~\eqref{eqg1d} can take very large values in a broad range of magnetic field, making the term in the denominator large and reducing the value of $g_{1D}$ even for considerably large scattering lengths.

\begin{figure}
  \centering
  \includegraphics[width=0.5\textwidth]{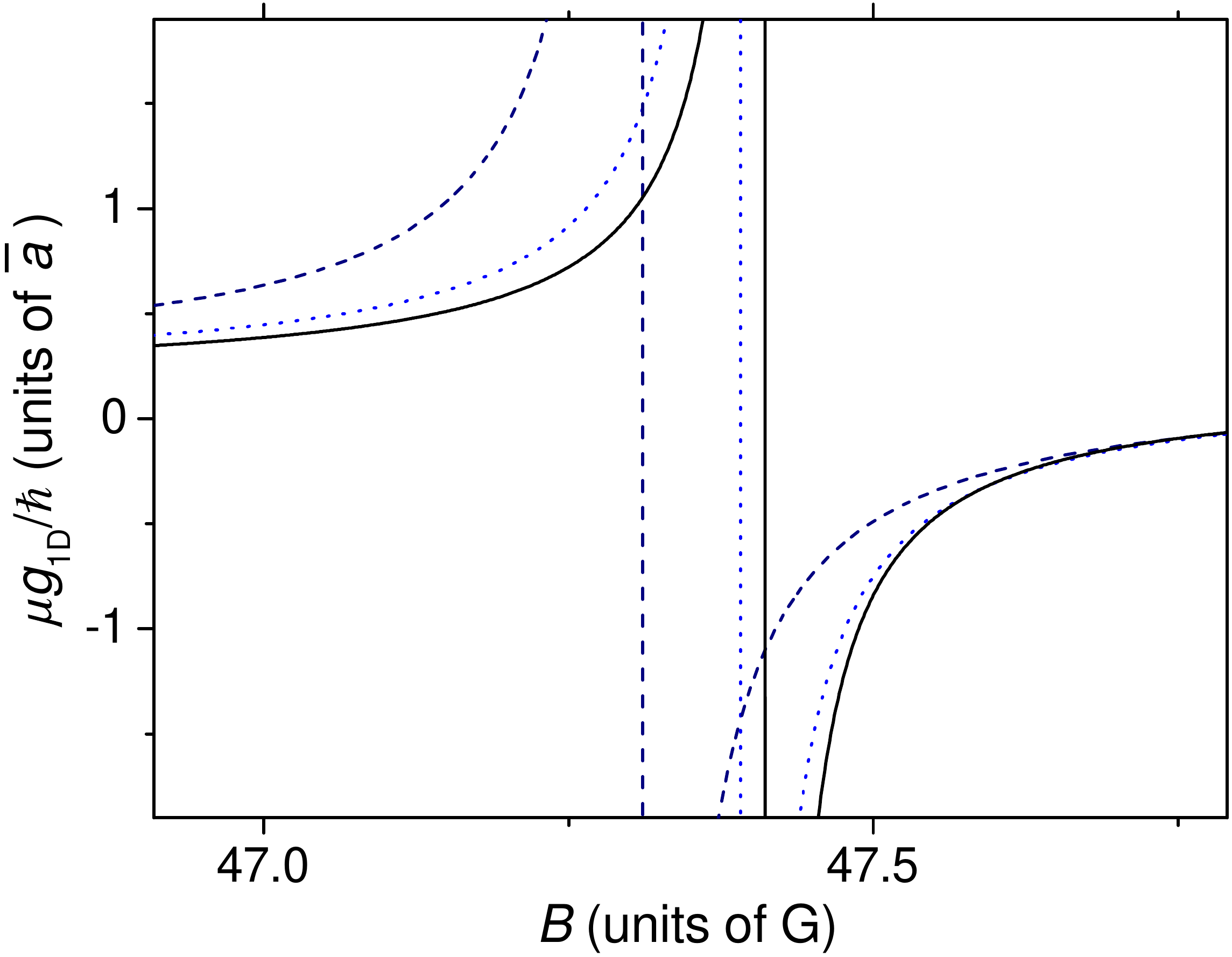}
  \caption{\label{fig:haller}The $g_{1D}$ coefficient described by eq.~\eqref{g1ddef} in the vicinity of the narrow $47.8$G resonance for $d=20\bar{a}$. The black solid line neglects the energy-dependent correction, whereas the dark blue dashed line is the full result assuming $p=0.01\bar{a}^{-1}$, and the light blue dotted line assumes $p=0.04\bar{a}^{-1}$ (which corresponds to around 220 nK).}
  \end{figure}

The $47.8$G resonance was used to study the CIR phenomenon~\cite{Haller2010}. Its position has been identified via increased three-body loss close to the pole of $g_{1D}$. A shift of the loss maximum to smaller values of $a_{3D}$ was identified when increasing the confinement strength. Specifically, $d$ was scanned in a range $19.9 \lesssim d/\bar{a} \lesssim 22.2$ leading to a well detectable shift of the loss feature by $\approx 100 \, \rm{mG}$ in magnetic field, which was found in agreement with the simplest formula~\eqref{a1D0}. The predicted shift due to the finite range correction shown in Figure~\ref{fig:haller} is of the same order for reasonable temperatures. However, the non-trivial shape of the experimentally observed loss resonance, which lacks a detailed theoretical model, makes it difficult to unambiguously identify the absolute position of the pole of $g_{1D}$. 

\begin{figure}
  \centering
  \includegraphics[width=0.5\textwidth]{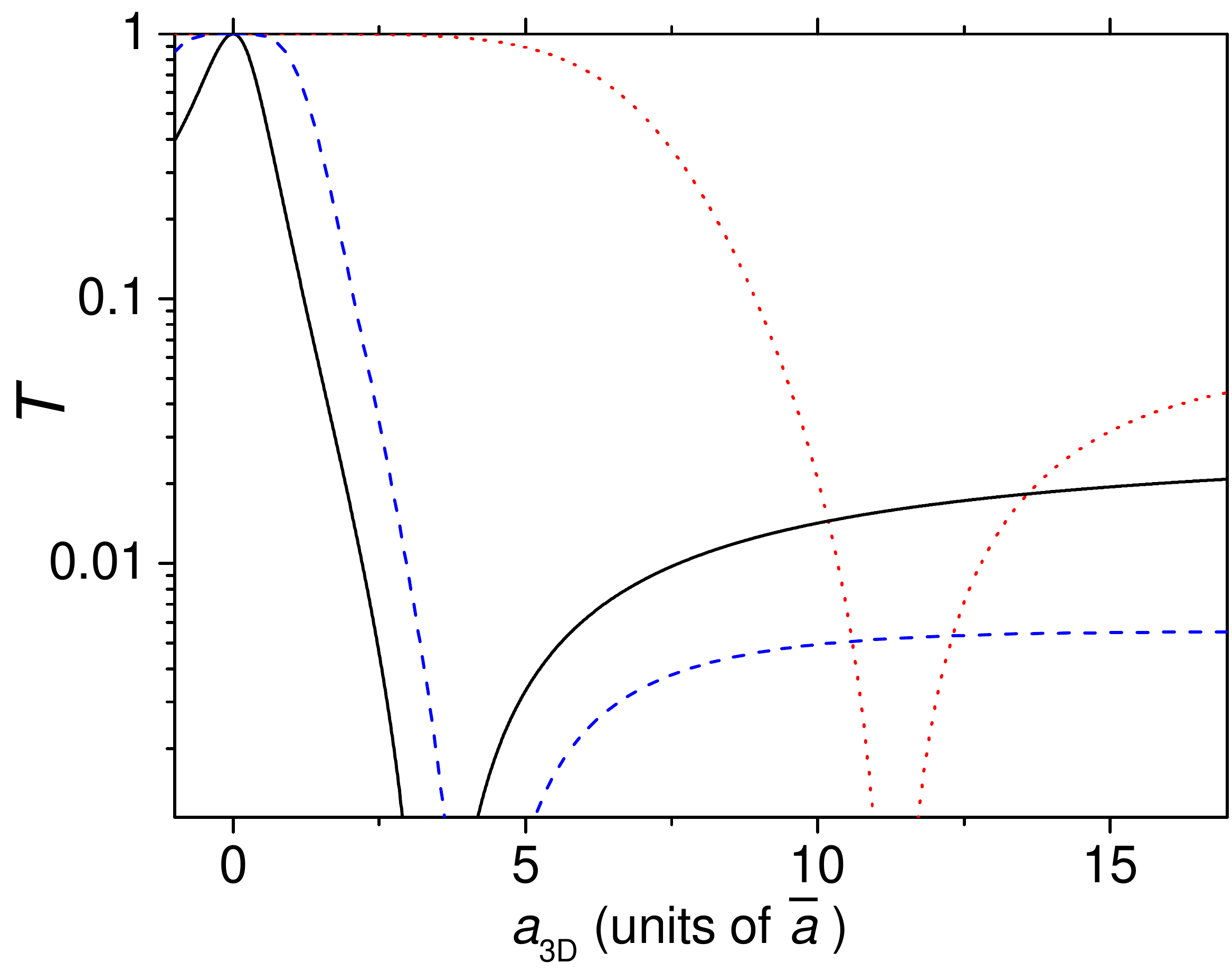}
  \caption{\label{fig:transmnarrow}Transmission coefficients for different Cs resonances (the black solid line for a broad resonance, the blue dashed line for the narrow $47.8$G resonance and the red dotted line for the narrow $226.7$G resonance) as in Fig.~\ref{fig:transmwide}, but for a narrow trap $d=5\bar{a}$ and with $p=0.02\bar{a}^{-1}$.}
\end{figure}

\begin{figure}
  \centering
  \includegraphics[width=0.5\textwidth]{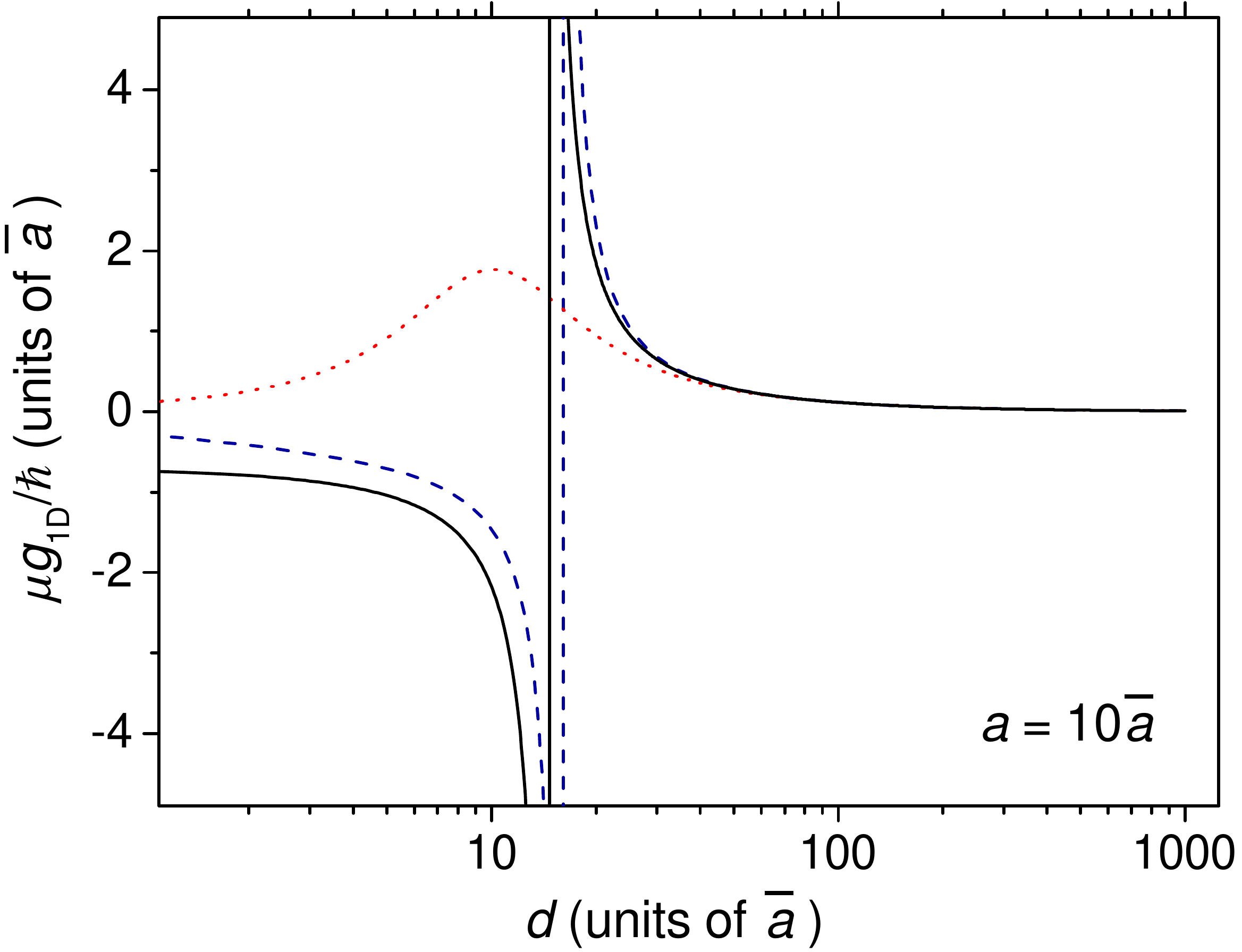}
  \caption{\label{fig:gd}$g_{1D}$ coefficient described by eq.~\eqref{eqg1d} as a function of the trap width for $a_{3D}=10\bar{a}$ for a broad resonance (black solid line), the narrow $47.8$G resonance (blue dashed line), and the resonance at $226.7$G (red dotted line).}
  \end{figure}

Observing large shifts in the position of CIR and magnitude of $g_{1D}$ would require going to stronger transverse confinement. In order to achieve $d/\bar{a} \sim 10$ it would be beneficial to use an optical lattice with shorter wavelength, e.g. 532 nm instead of 1064 nm. This is demonstrated in Figure~\ref{fig:transmnarrow}, which shows that for $d=5\bar{a}$ the position of the CIR related to the $226.7$ resonance is moved to much higher values of the 3D scattering length. However, the position of the CIR associated with the $47.8$G resonance is still not affected significantly. Figure~\ref{fig:gd} shows that the two narrow resonances display completely different behavior when the confinement strength is varied and the scattering length is kept constant.

The energy dependence of the resonance position can be readily investigated experimentally by varying the temperature of the trapped sample in a controlled way.
Alternatively, one may study the CIR of a single impurity interacting with the 1D gas, e.g. by using different Zeeman substates~\cite{Meinert2016}. As the impurity is accelerated by gravity or some other external force, its interaction with the background will change due to energy dependence and it might be possible to detect the modified transmission coefficient.

As has been shown in this section, effective range corrections and energy dependence are most pronounced in the vicinity of very narrow resonances and in tightly confined systems. In previous works the many-body properties of the quasi-1D gas were studied at moderate confinement strengths using a rather broad resonance, when for all practical purposes one could safely omit the energy dependence in the analysis of experimental results.

%\section{Implications for many-body physics}
%Let us now briefly discuss the impact of the modified pseudopotential for the many-body %physics. Changing the interaction from~\eqref{delta}, which is constant in momentum %space, to the form~\eqref{truepot} cannot affect the fundamental properties of the %system, such as the formation of Tonks-Girardeau gas at strong interactions. However, one %can expect some modification of the observable quantities such as momentum distribution %or the structure factor. It remains to be seen whether an exact solution of the %model~\eqref{truepot} can be developed.

\section{Relation to the generalized Lieb-Liniger model}

\begin{figure}
  \centering
  \includegraphics[width=0.5\textwidth]{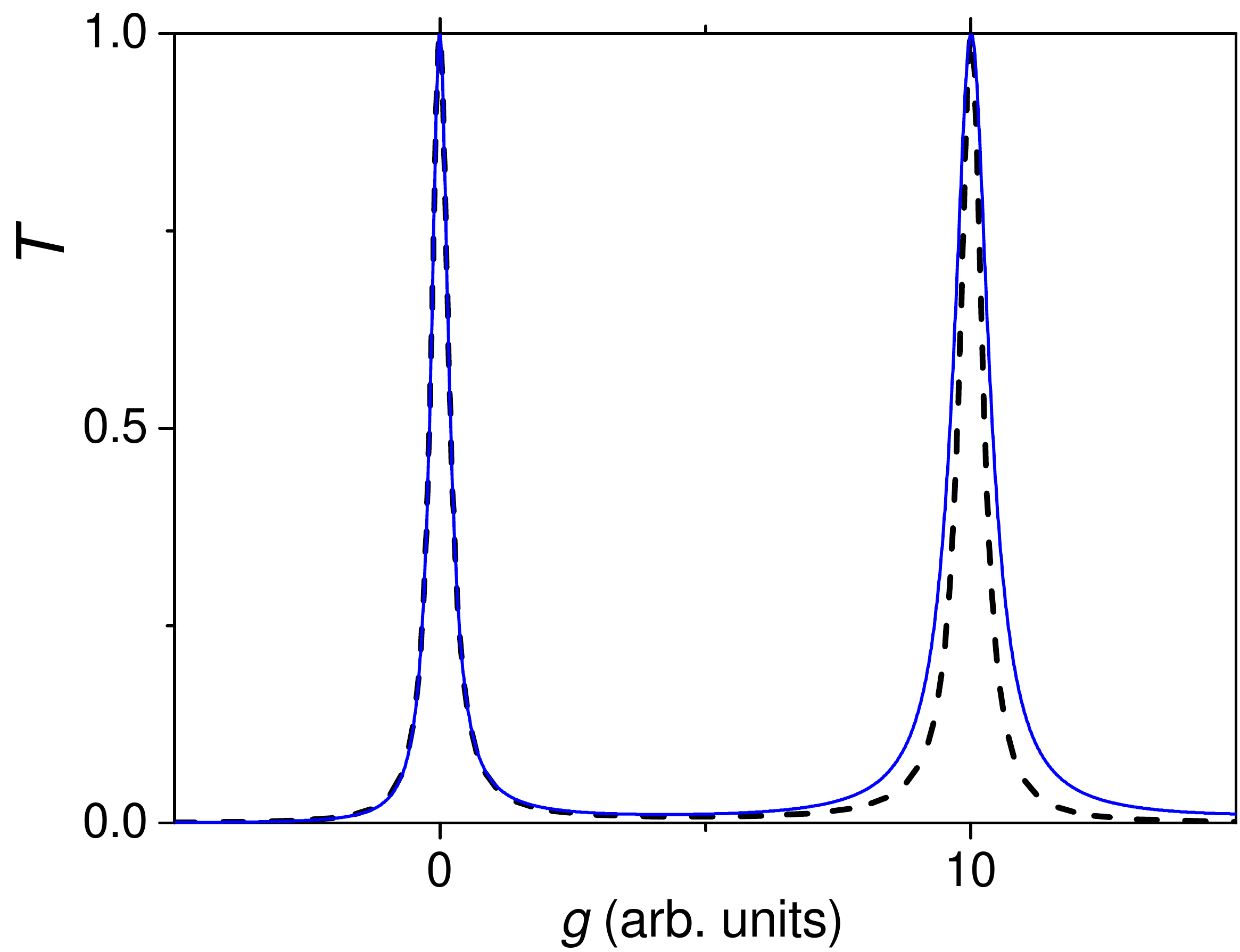}
  \caption{\label{fig:gll}Transmission coefficient calculated using interaction potential~\eqref{gll} (blue solid line) compared to the prediction of approximate result~\eqref{ceff} (black dashed line). Dimensionless units $\hbar^2/2\mu=1$ are used with energy $E=0.01$ and $\ell=0.1$. }
\end{figure}

\begin{figure}
  \centering
  \includegraphics[width=0.5\textwidth]{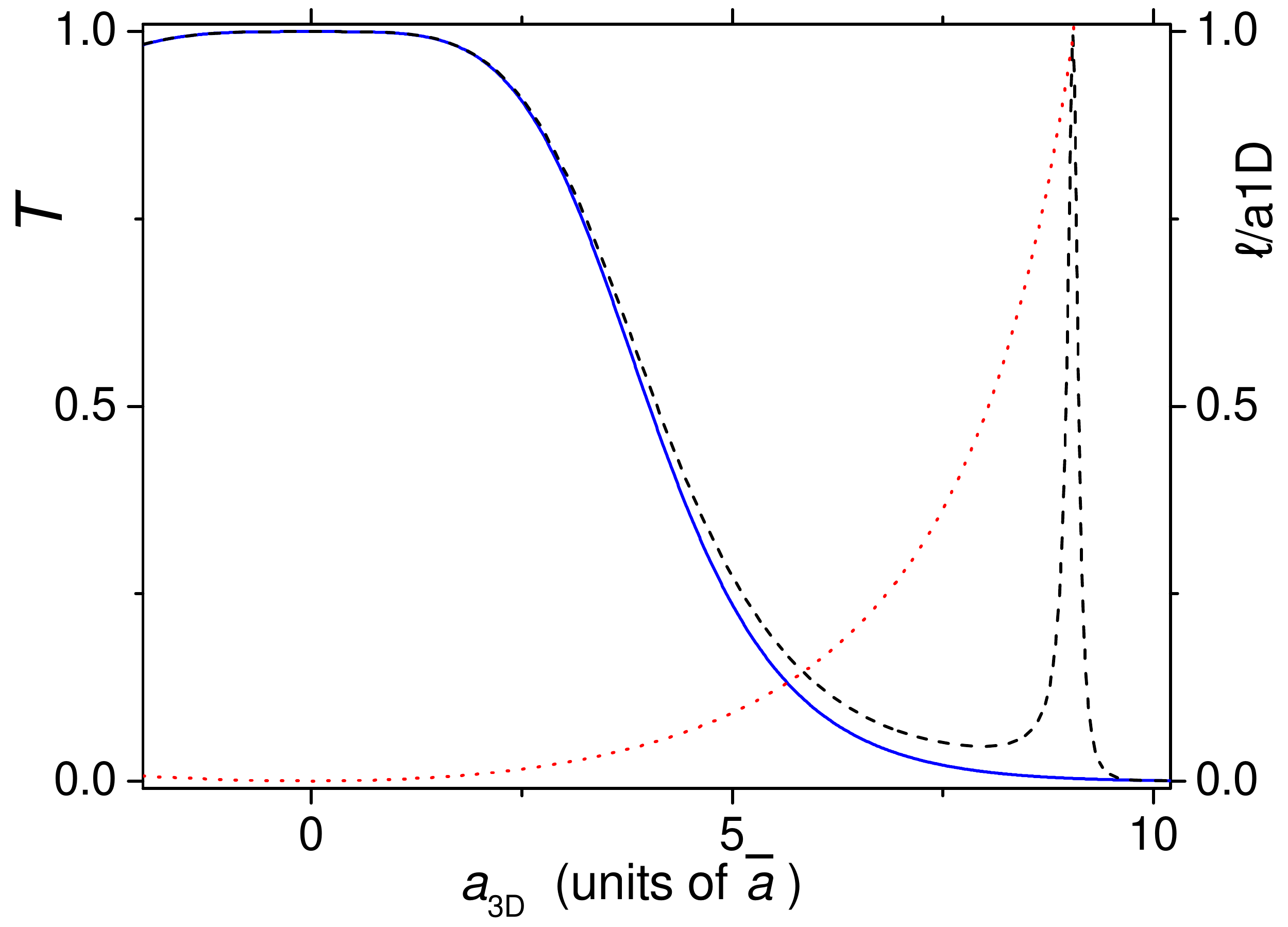}
  \includegraphics[width=0.45\textwidth]{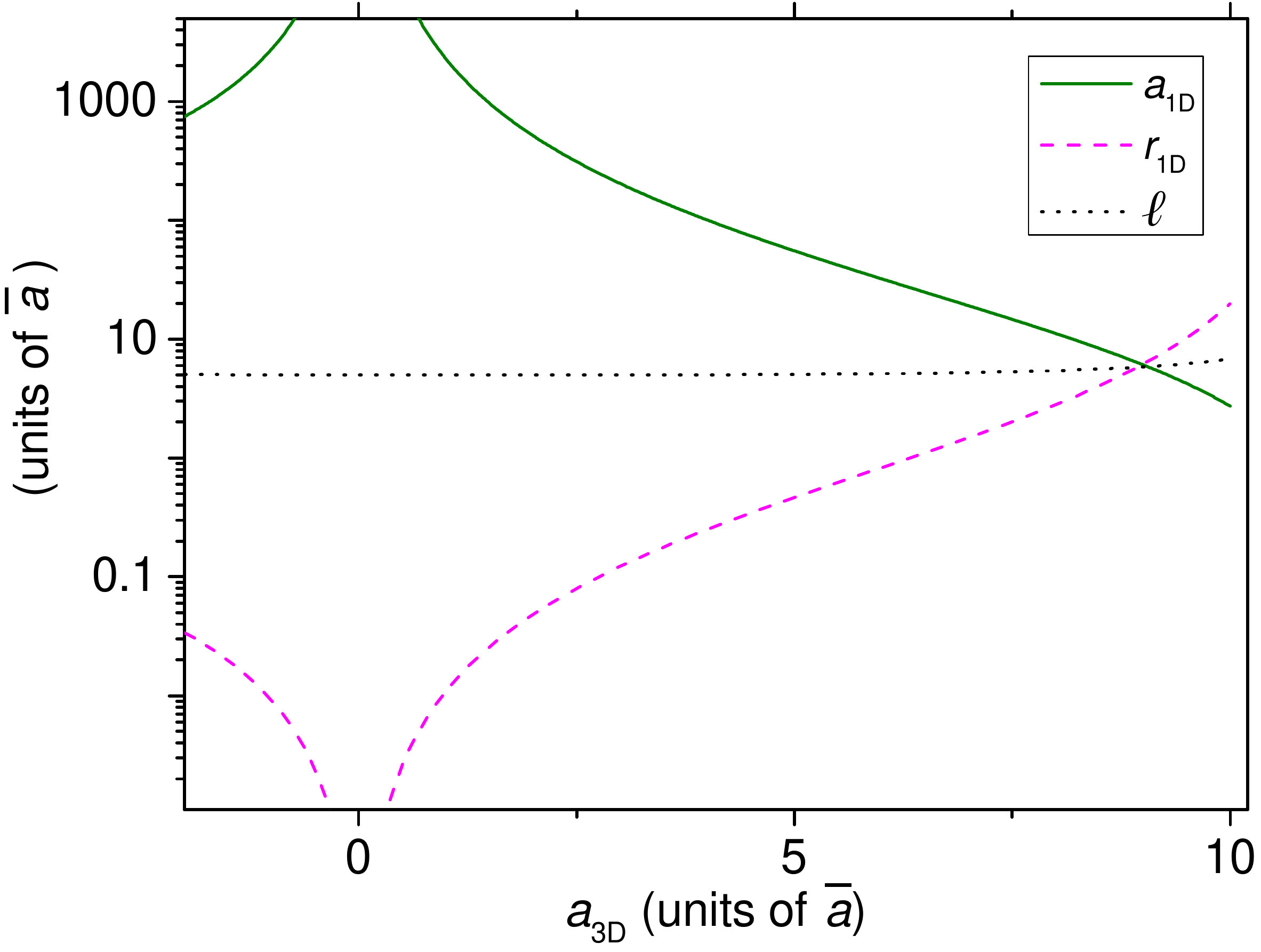}
\caption{\label{fig:gllcomp}Upper panel: transmission coefficient for the narrow Cs $226.7$G resonance (blue solid line) calculated using~\eqref{a1def},~\eqref{g1ddef}-\eqref{eqtransm} compared with the full  solution of the discretized potential~\eqref{eqgll} (blacked dashed line), which predicts a spurious unit transmission peak. Additionally, the red dotted line shows the ratio $\ell/a_{1D}$. The applicability of GLL model is limited to $\ell/a_{1D}\ll 1$. Here $d=5\bar{a}$ and $p=0.01\bar{a}^{-1}$, so that $p\ell\approx 0.05$ ($\ell$ is a weakly varying function of $a_{3D}$). Lower panel: behaviour of different length scales ($a_{1D}$, $r_{1D}$ and $\ell$) for the same resonance.} %new axis
\end{figure}

As shown in the previous section, the 1D energy shift resulting from interactions between two particles can be written as $\Delta E=g_{1D}\left(1+g^\prime p^2\right))\left|\psi(0)\right|^2$.
The correction is proportional to $p^2$, which in real space becomes the second derivative: $p=-i\frac{d}{dx}$. Thus, in real space we can write the effective pseudopotential in the following form
\beq
V(x)=g_{1D}\left[\delta(x)+\frac{g^\prime}{2}\left(\overleftarrow{\partial^2_x} \delta(x)+\delta(x)\overrightarrow{\partial^2_x}\right)\right].
\eeq
Connection to the generalized Lieb-Liniger model can be made by discretizing the second derivative, which leads to the interaction potential~\eqref{gll}~\cite{Veksler2014,Veksler2016}. However, there are two possible results for $c_0$ and $c_\ell$ parameters. When $g^\prime>0$, one can define a new length scale $\ell=\sqrt{2g^\prime}$. Then after discretization the pseudopotential becomes 
\beq
V(x)\approx g_{1D}\left(2\delta(x)-\frac{1}{2}(\delta(x+\ell)+\delta(x-\ell)\right).
\label{eqgll}
\eeq
In the case $g^\prime<0$, we instead have $\ell=\sqrt{-2g^\prime}$,
%\beq
%V(x)\psi(x)\approx g_{1D}\delta(x)+\frac{1}{2}g_{1D} \ell^2 \delta(x)\left(\frac{\psi(x+\ell)-2\psi(x)+\psi(x-\ell)}{\ell^2}\right),
%\eeq
which then gives
\beq
V(x)=\frac{g_{1D}}{2}(\delta(x+\ell)+\delta(x-\ell)).
\eeq
For $\ell\to 0$ both potentials reduce to the standard result~\eqref{delta} with strength $g_{1D}$. By comparing the results with~\eqref{gll}, we find
that for $g^\prime>0$ we have $c_0=2g_{1D}$ and $c_{\ell}=-g_{1D}/2$, while for $g^\prime<0$ we obtain $c_0=0$ and $c_{\ell}=g_{1D}/2$. Equation~\eqref{ceff} becomes then
\beq
c_{\rm eff}(g^\prime>0)=g_{1D}\left(1-\frac{q^2+3q}{2+q+q^2}\right)
\label{ceff2}
\eeq
and
\beq
c_{\rm eff}(g^\prime<0)=g_{1D}\left(1+\frac{q}{2-q}\right),
\eeq
where we have introduced $q=mg_{1D}\ell/\hbar^2$. The validity of the approximation utilizing $c_{\rm eff}$ given by~\eqref{ceff} compared to the actual potential~\eqref{gll} is shown in Figure~\ref{fig:gll}. However, the discretization procedure can only be justified if the introduced length scale $\ell$ is small enough. As the 1D scattering length is connected to $g_{1D}$ via~\eqref{g1ddef}, we have $q=2\ell/a_{1D}$. Consequently, one can expect that the predictions of GLL are valid for cold atoms only at $q\ll 1$. This conjecture is confirmed by a direct comparison of the GLL results with the full solution~\eqref{g1ddef} given in Figure~\ref{fig:gllcomp}, which shows that the agreement is lost exactly when the value of $q$ starts being comparable to one. In order to check that this is the only necessary condition of GLL applicability, we also investigated the behavior of the 1D effective range $r_{1D}$, which is given by expanding $1/a_{1D}(p)$ in $p$ in an analogous manner as done for the 3D case~\eqref{eqeffrexp}. The exemplary behaviour of all effective 1D length scales is shown in the lower panel of Fig.~\ref{fig:gllcomp}. We find that $r_{1D}$ is much smaller than $\ell$ for small scattering lengths where $q\ll 1$ and GLL is a good approximation for real systems. It grows considerably close to the CIR. Interestingly, all involved length scales $a_{1D}$, $r_{1D}$ and $\ell$ become comparable at the position of the spurious unit transmisison peak predicted by GLL (dashed black line in Fig.~\ref{fig:gllcomp}).

\section{Discussion and conclusions}
We have investigated how the atomic interactions in quasi-1D waveguides are affected by their realistic scattering properties. We found that the corrections manifest by the shift of the confinement-induced resonance and that the energy dependence of the effective interaction strength is especially important in the vicinity of narrow Feshbach resonances, and we demonstrated that using the example of Cs atoms. Although the strength of the corrections can differ depending on the particular resonance, they can still be cast in a universal analytic form using effective range expansion, as summarized by formulas~\eqref{g1ddef}-\eqref{eqgp}. In the future it would be useful to extend the analysis by setting up a truly multichannel model to take into account subtle effects such as differential polarizabilities of the bound states which may lead to further shifts in the resonance positions in the presence of the trapping light, and to study parameter regimes in which the effective range expansion is no longer sufficient.

The applicability of the generalized Lieb-Liniger model to the case of ultracold trapped atoms turned out to be limited. The necessary condition for this was determined to be $\ell/a_{1D}\ll 1$, otherwise the discretization of the second derivative is not justified. However, despite the lack of quantitative agreement, the prediction of a strong dependence of the interactions on the effective range is indeed confirmed by realistic calculations. In both cases we have found regions of effective transparency, although the origin of this effect is different in each case. The GLL thus served as a useful simple model and as a motivation to perform the more detailed analysis.

It would be interesting to look for systems for which the GLL model would provide a quantitatively correct description in the regime of unit transmission (where $c_{\rm eff}=0$, although $g_{1D}\neq 0$). This might be the case for more complex (e.g. molecular) systems.

In this paper we focused on scattering. An interesting future
direction may be to consider the impact of finite energy corrections on the thermodynamics of the many-body system. A potentially very interesting
regime for experimental studies is the case with $g_{1D}<0$. For the
narrow Cs resonance that we studied here, it is possible to realize
the case in which the effective 1D interactions are attractive
although $a_{3D}>0$, as one can see from Figs.~3 and~5. If the attraction  is
sufficiently weak (1D scattering length is large and negative), the
magnitude of $q$ may be much smaller than $1$ so that that the
approximation of the real system by the GLL holds. In  such a
situation, the ordinary Lieb-Liniger model~\eqref{delta}
with an effective strength $c_{\rm eff}$ that turns out to be negative provides a good description of the interactions.
It has been shown rigorously~\cite{Lieb1963} that in this case the ground
state energy scales with the number of particles as $-N^2$,  so
there is no thermodynamic limit and the system might be expected to
collapse. Then the dynamics cannot be described by the Gross-Pitaevskii equation. Since numerical calculations of the dynamics are generally limited to a small number of particles, this issue
should be tested by experiments. 

We would like to thank Immanuel Bloch for inspiring this collaboration and Panagiotis Giannakeas and Doron Cohen for fruitful discussions.
This work was supported in part by the Alexander von Humboldt Foundation, Polish National Science Center project 2014/14/M/ST2/00015, Israel Science Foundation (ISF) grants 1028/12 and 931/16, US-Israel Binational Science Foundation (BSF) grant number 2010132, Shlomo Kaplansky academic chair and by the US National Science Foundation under Grant No. NSF PHY-1125915. SF, KJ and PSJ are grateful for the hospitality of the Kavli Institute for Theoretical Physics in Santa Barbara, where part of this research was conducted. F.M. acknowledges support from Deutsche Forschungsgemeinschaft (DFG) within the SFB/TRR21 and the project PF 381/13-1.

\bibliography{Allrefs}
\end{document}